# Towards Simulating Social Influence Dynamics with LLM-based Multi-agents

Hsien-Tsung Lin
*Department of Information Management*
*National Sun Yat-Sen University*
Kaohsiung, Taiwan
aidenlin@g-mail.nsysu.edu.tw

Pei-Cing Huang
*Department of Information Management*
*National Sun Yat-Sen University*
Kaohsiung, Taiwan
pcpeicing@gmail.com

Chan-Tung Ku
*Department of Information Management*
*National Sun Yat-Sen University*
Kaohsiung, Taiwan
kuchantung@gmail.com

Chan Hsu
*Department of Information Management*
*National Sun Yat-Sen University*
Kaohsiung, Taiwan
chanshsu@gmail.com

Pei-Xuan Shieh
*Department of Information Management*
*National Sun Yat-Sen University*
Kaohsiung, Taiwan
iamshiehpay@gmail.com

Yihuang Kang
*Department of Information Management*
*National Sun Yat-Sen University*
Kaohsiung, Taiwan
ykang@mis.nsysu.edu.tw

*Abstract*—Recent advancements in Large Language Models offer promising capabilities to simulate complex human social interactions. We investigate whether LLM-based multi-agent simulations can reproduce core human social dynamics observed in online forums. We evaluate conformity dynamics, group polarization, and fragmentation across different model scales and reasoning capabilities using a structured simulation framework. Our findings indicate that smaller models exhibit higher conformity rates, whereas models optimized for reasoning are more resistant to social influence.

## I. Introduction

Recent advancements in machine learning, particularly in Large Language Models (LLMs), have substantially enhanced the capability of machines to emulate human language patterns, cognitive processes, and interactive behaviors [1]. By training on extensive datasets harvested from human-written texts across various platforms, these models implicitly capture linguistic structures, inherent human behavioral traits, and social dynamics [2]. Consequently, it is reasonable to hypothesize that sophisticated LLMs might exhibit human-like social behaviors [3] when engaged in interactive communication tasks. This prompts the question: can multiple LLM-based agents authentically mimic the social phenomena observed in human group interactions, including conformity, polarization, and fragmentation?

Online forums, specifically Bulletin Board Systems (BBS), represent ideal testbeds for evaluating such questions. A BBS typically fosters rich, asynchronous text-based discussions among multiple participants who mutually influence each other's opinions through repeated interactions. Participants in these platforms frequently display diverse social behaviors, dynamically adjusting their stances, forming consensus, or becoming increasingly polarized through dialogue. Leveraging LLMs within this context enables researchers to simulate and analyze intricate human interaction patterns at scale, addressing the traditional limitations associated with human-based experimental methodologies, such as participant scalability, ethical constraints, and resource intensity.

Given this promising scenario, the primary objective of our research is to explore the extent to which multi-agent simulations [4], driven by advanced LLMs, capture and reproduce authentic human social dynamics in forum-style environments. Specifically, this study addresses three pivotal research questions:

1. Social influence dynamics: To what extent do multi-agent dialogues among LLM-based agents replicate documented human social behaviors—particularly group polarization, conformity to majority viewpoints, and the fragmentation of collective opinions?

2. Model capacity: How does varying the parameter size of LLM-based agents affect the emergence and intensity of social influence phenomena, such as stance volatility and majority consensus?

3. Reasoning abilities: What role do specialized reasoning modules play in shaping each agent's susceptibility to peer pressure, stance retention, and overall polarization within the forum?

This paper offers several contributions to agentic artificial intelligence (AI) research and computational social science literature. We propose and validate a robust, structured multi-agent conversation framework explicitly designed to mimic the asynchronous interaction patterns typical of BBS forums. In addition, our empirical evaluation systematically investigates how different LLM capabilities and scales influence the dynamics of simulated social interactions, quantifying conformity rates, polarization tendencies, and opinion stability across multiple conversational rounds. Our analysis demonstrates the intricate interplay between model capabilities and social influence dynamics, underscoring the viability of our framework in capturing realistic interaction patterns inherent in human conversations [5].

The rest of this paper is structured as follows: In Section 2, we provide a quick review of relevant literature, covering key concepts in social simulations, LLM architectures, and prior research exploring human-like behaviors in LLM-based agents. In Section 3, we outline our proposed approach along with the multi-agent systems (MAS) setup, agent persona definitions, and evaluation metrics employed. In Section 4, we present experimental results and analyses, offering both quantitative and qualitative insights into the observed social phenomena. Finally, Section 5 summarizes our key findings and offers concluding remarks.

## II. BACKGROUND

Understanding the dynamics of group behavior and decision-making processes [6] is a longstanding research goal in fields ranging from social psychology and communication studies to computational social science. Traditional approaches to studying group interactions typically rely on controlled experiments, observational studies, or survey-based analyses. These methods, while valuable, encounter inherent limitations such as difficulties in scaling experiments to large groups, ethical constraints in manipulating certain interaction variables, and challenges in controlling environmental variables. To address these constraints, computational simulations leveraging agent-based models (ABM) have emerged as powerful alternatives, enabling researchers to systematically investigate complex social phenomena with precision, flexibility, and scalability [7].

Historically, ABM has been extensively utilized to simulate group interactions by defining simple behavioral rules for agents and observing emergent collective behaviors [8]. However, conventional ABM agents often lack realistic human behavioral complexity, primarily due to their oversimplified decision rules and deterministic interactions. The emergence of LLMs, such as OpenAI GPT-4 [9] and Google Gemini [10], has significantly expanded the potential of ABM by providing agents with advanced natural language processing capabilities [11], vast knowledge repositories, and nuanced communicative competencies. LLM-powered agents go beyond linguistic fluency, exhibiting behaviors such as adapting opinions to peer input, aligning with group norms, or intensifying initial attitudes.

LLMs, driven by transformer-based neural network architectures, excel in learning complex patterns directly from vast textual datasets encompassing diverse human-generated content. Training these models typically involves massive-scale unsupervised learning on internet-derived text corpora, embedding linguistic rules and capturing implicit social norms, cultural nuances, and conversational pragmatics. Therefore, LLM-based agents inherently possess features of human-like cognition, such as the ability to interpret context, reason logically, express attitudes, and adapt stances dynamically based on conversational cues and previous interactions. These attributes make LLMs promising candidates for simulating realistic human social behaviors, particularly within dynamic communication environments like online forums [12], [13].

BBS, characterized by asynchronous text-based communication among multiple users, offers a unique window into real-world social interaction dynamics. The inherent design of BBS forums facilitates prolonged and reflective exchanges where users iteratively express opinions, challenge others, defend viewpoints, and occasionally shift their stances based on group discussions. Previous research on online forum interactions has highlighted distinct social influence phenomena, including *conformity*, *group polarization*, and *fragmentation.* Conformity refers to individual adjustments of beliefs or behaviors to align with group norms. Group polarization describes the tendency for initial moderate positions to become more extreme through interaction [14], while fragmentation is characterized by divisions into subgroups holding opposing views [15], [16], [17]. Such social behaviors have significant implications across various contexts, including online governance, policy deliberation, marketing strategies, and public opinion management.

Integrating LLMs with MAS in simulating BBS-style discussions provides a substantial methodological advancement in computational social science. Recent studies have begun exploring this intersection. For instance, Gao et al. demonstrated that GPT-based agent networks could simulate social learning and cooperative behaviors in digital communities [18]. Similarly, Park et al. showed that LLM-based agents could convincingly emulate human-like interactions within simulated virtual environments, offering valuable insights into collective behaviors and social decision-making processes [19]. These findings underscore the feasibility and potential of employing LLM-based MAS frameworks to systematically examine complex social phenomena in virtualized yet realistic interaction settings.

However, despite promising initial results, critical gaps remain in our understanding. First, it is unclear whether LLM-generated conversation is linguistically and behaviorally indistinguishable from authentic human interactions in online forums, raising concerns for content moderation and ethical deployment. Second, empirical evidence is still lacking regarding how model architecture, reasoning ability, and parameter scale shape the realism and stability of agent behavior [4], [20]. Addressing these gaps is crucial for improving the fidelity of LLM-based simulations and guiding responsible deployment practices in real-world scenarios [21].

In this paper, we address these gaps by systematically evaluating how varying LLM capacities and architectures influence the social dynamics of simulated BBS discussions. By bridging theoretical insights from social psychology [22] with state-of-the-art developments in agentic AI, our research contributes to both the computational social sciences and AI literature, providing rigorous empirical findings and methodological guidance for future research.

## III. SIMULATING SOCIAL INFLUENCE DYNAMIC WITH LLM-BASED AGENTS

We here present our proposed approach, an LLM-based, multi-agent conversational environment that simulates human-like interactions on BBS-style platforms. As shown in Fig. 1, the system architecture features a central manager node that orchestrates a round-robin message exchange: each agent posts in a predetermined sequence, and all messages are broadcasted to every participant (agent). This design enables agents to reference and respond to prior posts as if reading a live forum thread.

Each agent is defined by a structured persona prompt encompassing demographic attributes, communicative style, and a fixed stance on the assigned topic. This design ensures that a given persona maintains the same baseline position across different LLMs [23]. For example, Role A is presented as an idealistic and receptive individual who consistently supports the proposal until strong counterarguments are presented. This standardization allows us to isolate the effects of model architecture on social influence and stance evolution. The text-based interactions proceed through five consecutive rounds of posting. At the beginning of Round 1, the manager announces a controversial question or topic, such as whether governments should adopt stringent environmental policies. Each agent then submits an initial statement that reflects its persona and stance. Subsequent rounds require agents to quote or reference earlier posts from the conversation log. This structure imposes asynchronous turn-taking, maintaining clarity and ensuring all agents have equal speaking opportunities. Once Round 5 concludes, every agent has published five posts, leaving a complete conversation trace that captures their individual stances, rhetorical transitions, and potential opinion changes.

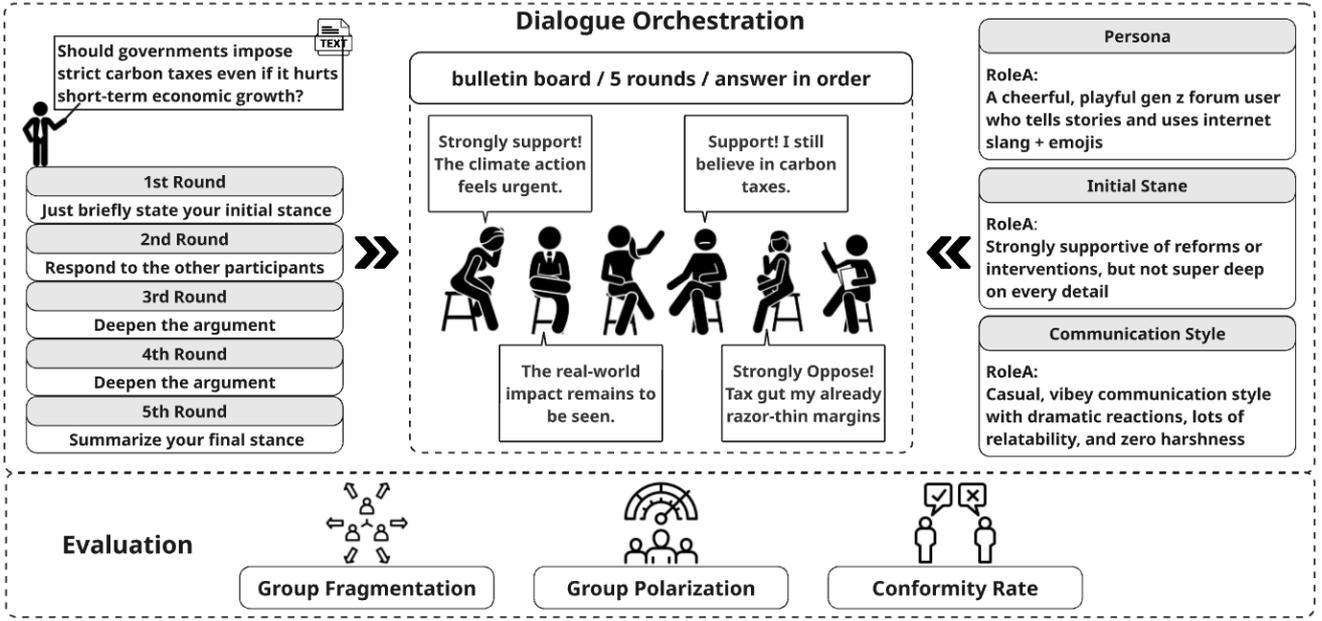

Fig. 1. Architecture for the Forum Simulation

To ensure the robustness of our findings in light of the inherent variability in LLM outputs, we treated each simulation setting as a complete five-round forum-style conversation and repeated it independently 25 times. Each repetition was considered a distinct simulation trial, encompassing the full sequence from Round 1 through Round 5 as previously described. We aggregated the results across these 25 trials, focusing on the overall patterns of conformity rate, polarization change, and fragmentation index. This procedure enables us to evaluate the stability and generalizability of the observed social behaviors under consistent conversational conditions.

We next would like to evaluate the aforementioned three main social phenomena: Conformity, Group Polarization, and Group Fragmentation. First, as discussed, Conformity in human group behavior typically arises when an individual adjusts a stance or opinion to align with the majority view [24]. In this simulation, each of the five agents declares an initial stance in Round 1 and has four additional opportunities to update or maintain that stance. Whenever an agent's shift in position brings it closer to the prevailing stance among the group. Such an event is considered a "conforming stance change." Let $N$ denote the total number of stance-change opportunities, which is the product of six agents, for instance, and four update windows, yielding $N = 24$. We define the Conformity Rate (CR) as follows:

$$CR = \frac{\sum_{i=1}^{N} \mathbf{1}\{\text{change } i \text{ aligns with majority}\}}{N} \quad (1)$$

where $\mathbf{1}\{\cdot\}$ is an indicator function that equals 1 if the $i^{\text{th}}$ stance change moves an agent's position toward the group majority, and 0 otherwise. A higher CR suggests that agents are more susceptible to perceived social pressure within the forum simulation. On the other hand, a lower CR indicates that agents prioritize their preexisting beliefs or are less influenced by others' statements.

Another crucial but common phenomena is Group Polarization, which refers to the tendency of a collection of individuals with varied initial viewpoints to converge on increasingly extreme stances over time. In this study, each agent's stance is represented on a five-point scale: Strongly Oppose (-2), Oppose (-1), Neutral (0), Support (+1), and Strongly Support (+2). After each discussion round concludes, the most recent stance of every agent is recorded, enabling us to track shifts in the overall stance distribution from the center of the scale toward either extreme. To further quantify the degree of polarization, we treat each round's stance distribution as a probability vector. Specifically, let $p_r(s)$ be the fraction of agents holding a stance $s$ at round $r$. Our Polarization Index $P_r$ at round $r$ is the expected absolute stance:

$$P_r = \sum_{s=-2}^{+2} |s| \cdot p_r(s) \quad (2)$$

where $|s|$ ranges from 0 to 2, and thus $P_r$ is bounded between 0 (all agents at Neutral) and 2 (all agents at an extreme). For instance, if the stance distribution in a given round is $\{p_r(-2) = 0.166, p_r(-1) = 0.166, p_r(0) = 0.336, p_r(+1) = 0.166, p_r(+2) = 0.166\}$, then $P_r = (2 \times 0.166) + (1 \times 0.166) + (0 \times 0.336) + (1 \times 0.166) + (2 \times 0.166) = 0.83$, which naturally falls near the midpoint of the 0–2 spectrum. As a result, it reflects a moderate spread of opinions ranging from Strongly Oppose (-2) to Strongly Support (+2).

Since polarization in a multi-round simulation reflects how stances evolve, we also consider the variation in $P_r$ from the initial round $r = 1$ through the final round $r = 5$. Specifically, we define a Polarization Change $\Delta P$ as: $\Delta P = [P_5 - P_1]$. These metrics help determine whether a given model exhibits increasing or stable polarization over multiple discussion rounds. For example, if a model starts at $P_1 = 0.83$ and ends at $P_5 = 1.53$, the $\Delta P = 0.7$ signifies a meaningful shift toward more extreme stances. This is especially significant when typical fluctuations are lower in other model scenarios.

Lastly, Group Fragmentation describes the condition in which conversation participants split into distinct subgroups holding fundamentally opposing positions, rather than converging on a unified consensus by the final round. In this setting, fragmentation is determined by the presence of two dominant stance clusters: one substantially favoring the proposition (Support or Strongly Support) and another strongly opposing it (Oppose or Strongly Oppose). Neutral stances, if present, remain relatively small in number and do not bridge the divide.

To capture the intensity of this division, we define a Fragmentation Index $F_r$ in terms of the proportions of agents supporting or opposing the proposition at round $r$:

$$F_r = 1 - \frac{|[p_r(+1)+p_r(+2)]-[p_r(-2)+p_r(-1)]|}{[p_r(+1)+p_r(+2)]+[p_r(-2)+p_r(-1)]} \quad (3)$$

Here, $p_5(+1) + p_5(+2)$ is the proportion of agents supporting the proposition, and $p_5(-1) + p_5(-2)$ is the proportion of agents opposing it, both measured in the final round. Higher values of $F_5$ imply a more balanced split between supporters and opponents, thus more substantial fragmentation. For example, if 20% of agents strongly support and 20% strongly oppose (with 60% neutral), approaches 1, indicating near-equal clusters at the extremes; conversely, if nearly all agents converge on one side, $F_5$ diminishes toward 0, reflecting minimal fragmentation.

Our multi-agent design, round-based posting framework, and explicit persona prompts offer an environment where conversation and behavioral metrics can be rigorously measured. The following sections detail how various LLM configurations were tested under this methodological framework and present the results of these experiments, including both numerical comparisons of $\Delta P$ and $F_5$ and qualitative examinations of evolving stance distributions.

## IV. EXPERIMENT AND DISCUSSION

Our experiments were conducted under uniform conditions, following the methodology described in Section III, so that each LLM faced the same discussion environment and topic prompts. These multi-agent simulations were implemented in Microsoft AutoGen [25], ensuring consistent orchestration of persona-based interactions and message exchanges across different model groups. We distinguished four principal groups of models according to their parameter scales, computational resource requirements, and inherent reasoning features. When exposed to identical experimental setups, this classification examined how varying model capacities and architectures affect social behaviors, such as conformity and stance evolution.

To enable robust cross-model comparisons, we organized the selected LLMs into the abovementioned four groups based on their computational requirements and architectural emphasis.

1. Group A: Operable on a single GPU, thus balancing accessibility with linguistic competence (i.e., Qwen2.5-7B [26], Llama3.1-7B [27], and Deepseek-R1-8B [28]

2. Group B: Offering higher capacity while remaining feasible for environments with limited computing resources (i.e., Qwen2.5-72b [26], Llama3.1-70B [27], and Deepseek-R1-70B [28]).

3. Group C: Represents widely adopted proprietary LLMs such as GPT-4o [29], Claude 3.5 Haiku[30], and Gemini Flash 2.0 [10].

4. Group D: Consists of architectures explicitly designed or fine-tuned for logical inference and reasoning, such as GPT-o1-mini [31], Deepseek-R1 [28], and QwQ-32B [32].

By maintaining a unified experimental design and consistent persona assignments across all groups, we control for external confounds, which enables a clear attribution of observed behavioral differences to model-specific factors. This setup allows for systematic analysis of how parameter scale, computational constraints, and reasoning orientation jointly shape individual agent stances and emergent group-level dynamics.

To evaluate the influence of model architecture on social alignment, we analyzed CRs across the four groups, as shown in Fig. 2. Most models in Groups A, B, and C fall within the 10–20% range, suggesting moderate responsiveness to peer influence. Among them, ChatGPT-4o reached the highest CR at 19.45%, implying that larger generative models may be more prone to majority alignment. By contrast, reasoning-oriented models in Group D showed considerably lower CRs, with ChatGPT-o1-mini at just 3.13%, suggesting a more substantial capacity to retain initial viewpoints under social pressure, possibly due to more consistent internal reasoning processes.

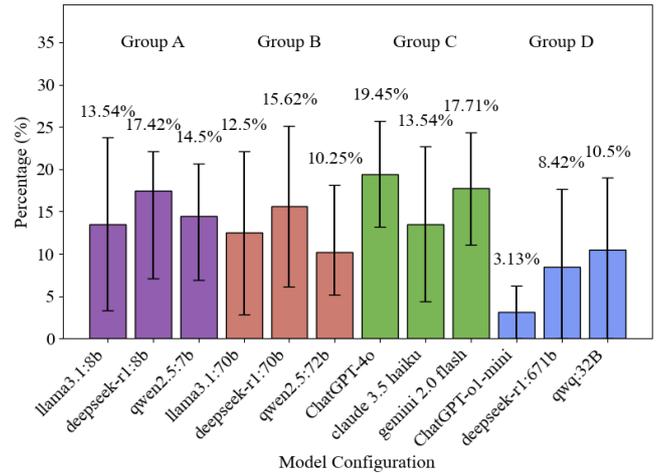

Fig. 2. Conformity Rates Among Four Model Groups.

Beyond static conformity, we examine how stance distributions evolve. As shown in Fig. 3, Groups A and B, have relatively higher $\Delta P$ and lower $F_5$ than Groups C and D, which suggests greater openness of Groups A and B to external influence and an increasing preference for "support" or "strongly support" $p_5(+1) + p_5(+2)$ over the rounds. This pattern also indicates that smaller or mid-sized models may lean toward consensus when their reasoning capabilities are limited. Notably, qwen2.5-72b finishes markedly higher $F_5 = 0.74$ than other models in the same group, indicating that it retains strong opposing factions even as overall stances shift. A similar pattern emerges in Group A's qwen2.5-7b, which achieves $F_5 = 0.33$, still higher than its immediate peers, further suggesting that the Qwen architecture, whether smaller or mid-sized, can preserve dissent under certain conditions.

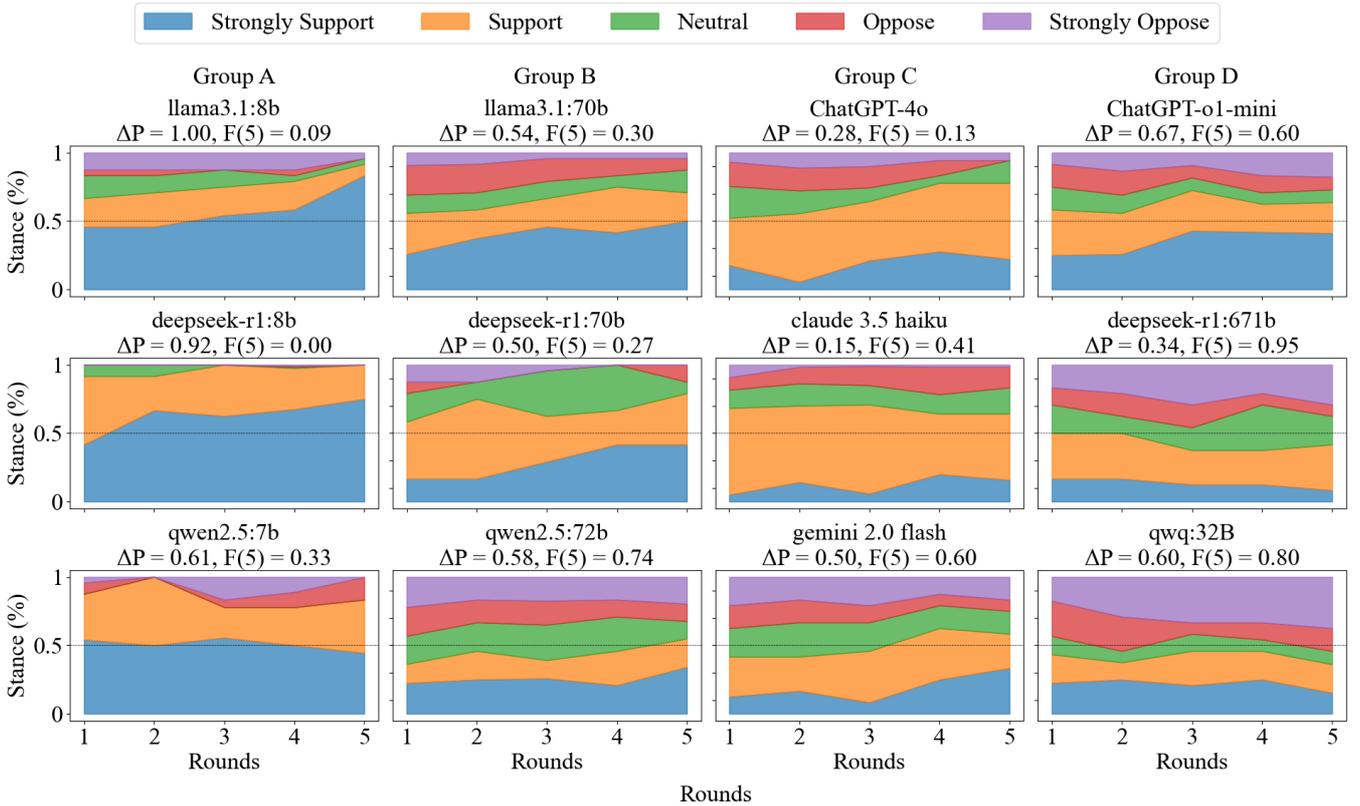

Fig. 3. Proportions of Agent Stances Over Successive Rounds.

Besides, Group C shows the lowest overall $\Delta P$, implying stronger resilience against extreme stance shifts. Within this group, ChatGPT-4o's $F_5 = 0.13$ signals that most agents ultimately adopt a supportive stance, diluting extreme opinions. Nevertheless, Group C's advanced architectures maintain more consistent viewpoints across rounds, as indicated by their reduced $\Delta P$. Finally, Group D consistently features a subset of agents in the "strongly oppose" category, suggesting that reasoning-focused models (or large reasoning models) can hold firm adversarial stances even when the broader conversation trends are supportive. Fragmentation also appears prominently in Group D; for instance, specific models exhibit a pronounced split between "strongly support" and "strongly oppose," signaling that logic-centric designs retain diverse opinions and allow dissenting views to persist alongside majority positions.

Our experimental results demonstrate that LLM-based multi-agent simulations can reproduce social phenomena—moderate conformity, group polarization, and persistent dissent. The relative stability and fragmentation observed in Group D (LRMs) suggest that models emphasizing logical consistency and structured reasoning may be better suited for applications requiring stance durability or viewpoint heterogeneity, such as deliberative AI agents or argumentation-based systems. By contrast, a mid-size or large generative model appears more prone to real-time alignment with the majority, particularly when repeated rounds of messaging cultivate a perceived consensus. From an application standpoint, these findings also suggest that researchers seeking to simulate extreme stance shifts or group consensus might consider LLMs with simple generative capacities. In contrast, those examining persistent dissent or vigorously defended positions might choose more reasoning-focused LLMs or LRMs. Ultimately, model selection should reflect the target phenomenon: do we want to observe realistic drift and consensus formation, or must we maintain heterogeneity and allow contrarian stances to flourish?

## V. CONCLUSION

We investigated how multi-agent conversations among LLM-based agents reflect human social influence phenomena, particularly those involving group polarization, conformity to majority viewpoints, and the fragmentation of collective opinions. We demonstrated that models with varying parameter sizes exhibit different levels of stance volatility and consensus formation: larger or mid-sized generative architectures tend to align with majority views, whereas models with specialized reasoning modules retain greater independence from external influence.

Our findings suggest that social influence dynamics emerge in LLM-based forums, mirroring documented human patterns of conformity and polarization. The model capacity shapes an agent's tendency to conform. However, larger size does not necessarily translate into either stronger or weaker susceptibility—rather, it interacts in complex ways with repeated rounds of consensus-building. Reasoning abilities serve as a critical buffer against majority pressure, enabling specific agents to preserve dissenting or adversarial positions. Therefore, model selection should align with research goals: those aiming to simulate realistic shifts toward consensus might choose simpler or mid-tier generative models, while investigations requiring persistent dissent or stance durability may benefit from more reasoning-focused architectures.